\documentclass[aps,prl,reprint,superscriptaddress,showpacs,showkeys]{revtex4-1}
\usepackage{amssymb}
\usepackage{amsmath}
\usepackage{cancel}
\usepackage{color}
\usepackage{braket}
\usepackage{graphicx}
\usepackage{epstopdf}
\usepackage{verbatim}
\usepackage{siunitx}
\usepackage{soul}
\usepackage{xr}
\usepackage[toc,page]{appendix}
\usepackage[section]{placeins}

\newcommand{\cnuc}{$^{\mathrm{13}}$C}

\begin{document}

\title{Quantum network nodes based on diamond qubits with an efficient nanophotonic interface
}

\author{C. T. Nguyen}
\email[These authors contributed equally to this work\\]{christiannguyen@g.harvard.edu}
\affiliation{Department of Physics, Harvard University, Cambridge, Massachusetts 02138, USA}

\author{D. D. Sukachev}
\email[These authors contributed equally to this work\\]{christiannguyen@g.harvard.edu}
\affiliation{Department of Physics, Harvard University, Cambridge, Massachusetts 02138, USA}

\author{M. K. Bhaskar}
\email[These authors contributed equally to this work\\]{christiannguyen@g.harvard.edu}
\affiliation{Department of Physics, Harvard University, Cambridge, Massachusetts 02138, USA}

\author{B. Machielse}
\email[These authors contributed equally to this work\\]{christiannguyen@g.harvard.edu}
\affiliation{Department of Physics, Harvard University, Cambridge, Massachusetts 02138, USA}
\affiliation{John A. Paulson School of Engineering and Applied Sciences, Harvard University, Cambridge, Massachusetts 02138, USA}

\author{D. S. Levonian}
\affiliation{Department of Physics, Harvard University, Cambridge, Massachusetts 02138, USA}

\author{E. N. Knall}
\affiliation{John A. Paulson School of Engineering and Applied Sciences, Harvard University, Cambridge, Massachusetts 02138, USA}

\author{P. Stroganov}
\affiliation{Department of Physics, Harvard University, Cambridge, Massachusetts 02138, USA}

\author{R. Riedinger}
\affiliation{Department of Physics, Harvard University, Cambridge, Massachusetts 02138, USA}

\author{H. Park}
\affiliation{Department of Physics, Harvard University, Cambridge, Massachusetts 02138, USA}
\affiliation{Department of Chemistry and Chemical Biology, Harvard University, Cambridge, Massachusetts 02138, USA}

\author{M. Lon\v{c}ar}
\affiliation{John A. Paulson School of Engineering and Applied Sciences, Harvard University, Cambridge, Massachusetts 02138, USA}

\author{M. D. Lukin}
\email{lukin@physics.harvard.edu}
\affiliation{Department of Physics, Harvard University, Cambridge, Massachusetts 02138, USA}

\begin{abstract}
Quantum networks require functional nodes consisting of stationary registers with the capability of high-fidelity quantum processing and storage, which efficiently interface with photons propagating in an optical fiber. 
We report a significant step towards realization of such nodes using a diamond nanocavity with an embedded silicon-vacancy (SiV) color center and a proximal nuclear spin. 
Specifically, we show that efficient SiV-cavity coupling (with cooperativity $C >30$) provides a nearly-deterministic interface between photons and the electron spin memory, featuring coherence times exceeding \SI{1}{\milli\second}. 
Employing coherent microwave control, we demonstrate heralded single photon storage in the long-lived spin memory as well as a universal control over a cavity-coupled two-qubit register consisting of a SiV and a proximal \cnuc~ nuclear spin with nearly second-long coherence time, laying the groundwork for implementing  quantum repeaters.

\end{abstract}

\maketitle

%%%% Introduction %%%%
The realization of quantum networks is one of the central challenges in 
quantum science and engineering with potential applications to long-distance communication, non-local  sensing and metrology, and distributed quantum computing \cite{kimble2008quantum,childress2005fault,gottesman2012sensing,komar2014quantum,monroe2014large}. 
Practical realizations of such networks require individual nodes with the ability to process and store quantum information in multi-qubit registers with long coherence times, and to efficiently interface these registers with optical photons.
Cavity quantum electrodynamics (QED) is a promising approach to enhance interactions between atomic quantum memories and photons \cite{Stute2012tunable,reiserer2015cavity,kalb2015heralded,lodahl2015interfacing,welte2018photon}. 
Trapped atoms in optical cavities are one of the most developed cavity QED platforms for quantum processing, and have demonstrated gates between atoms and photons \cite{reiserer2014quantum} as well as interactions between multiple qubits mediated by the optical cavity \cite{welte2017cavity}. 
While these experiments have demonstrated all of the individual components needed for a quantum network, combining them to realize a full-featured node remains an outstanding challenge. 

Nanophotonic cavity QED systems with solid-state emitters are appealing candidates for realizing quantum nodes as they can be interfaced with on-chip electronic control and photonic routing, making them suitable for integration into large-scale networks \cite{lodahl2015interfacing, Molesky2018inverse}.
Numerous advances towards the development of such nodes have been made recently. 
Self-assembled quantum dots in GaAs have been efficiently interfaced with nanophotonic structures, enabling a fast, on-chip spin-photon interface \cite{lodahl2015interfacing, sun2016quantum}. 
Nitrogen-vacancy color centers in diamond (NVs) have demonstrated multi-qubit quantum processors with coherence times approaching one minute \cite{bradley2019minute}, and have been used to implement quantum error correction \cite{Waldherr2014Quantum} and teleportation \cite{Pfaff2014unconditional}. 
Despite this rapid progress, functional nodes combining all the necessary ingredients in a single device have not yet been realized. 
For example, quantum memory times in quantum dots are limited to a few \si{\micro\second} by the dense bath of surrounding nuclear spins \cite{huthmacher2018coherence}. 
Conversely, an efficient nanophotonic interface to NVs remains elusive, in part due to the degradation of their optical properties inside nanostructures arising from electrical noise induced by the fabrication \cite{faraon2012coupling, ruf2019optical}. 

%%%%%%%%%%%%%%%%%% transition to technical part: summary
%
    \begin{figure}
		\includegraphics[width=\linewidth]{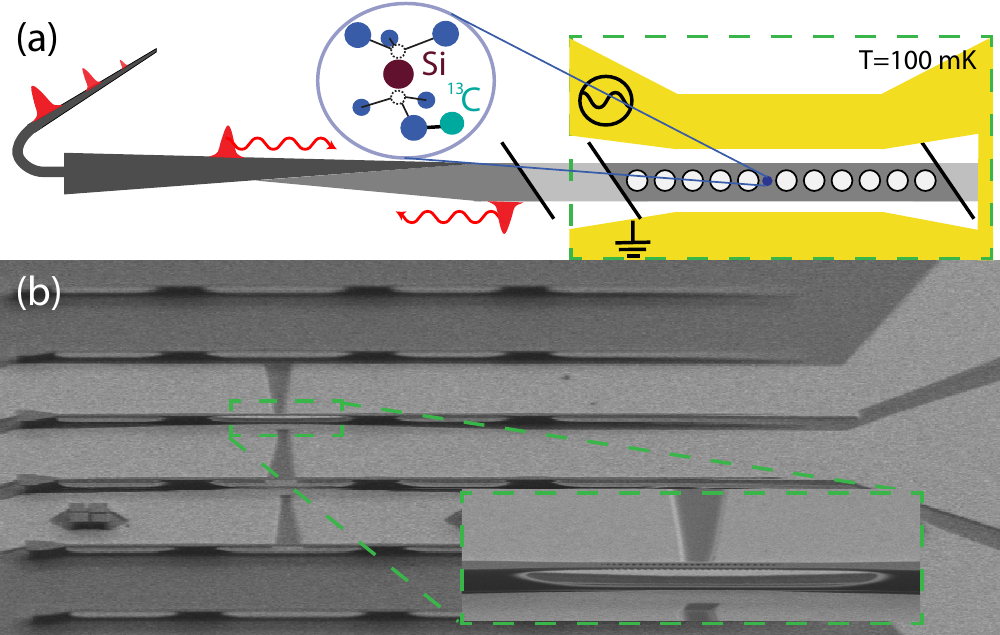}
	\caption{
			(a) Schematic of a SiV-nanophotonic quantum register. A diamond nanostructure with embedded SiV centers and ancillary \cnuc~nuclei are coupled via a waveguide to a fiber network. Spins are controlled by an on-chip microwave CPW at \SI{0.1}{\kelvin}. 
			%Photons are extracted into an optical fiber and can be distributed to the network.
			%\textbf{b.} Experimental setup. Silicon-vacancy centers are implanted into the mode maximum of nanophotonic crystal cavities. These devices are cooled to \SI{100}{\milli\kelvin} in a dilution refrigerator, and coupled to a tapered optical fiber. Resonant light probes the atom-cavity system, and is measured in reflection. A gold stripline patterned onto the diamond surface applies microwave pulses for controlling the SiV spin.
			(b) Scanning electron micrograph of several devices. The gold CPW is designed to localize microwave fields around the cavity center (green inset).
			%showing an array of diamond nanostructures addressed by a lithographically aligned gold microwave CPW. The CPW is designed to localize microwave fields around the cavity center (green inset). %Tapered diamond waveguides (blue inset) allow for efficient photon collection.
			}
	    \label{fig:network}
    \end{figure}
In this Letter, we demonstrate an integrated network node combining all key ingredients required for a scalable quantum network. 
This is achieved by coupling a negatively charged silicon-vacancy color-center (SiV) to a diamond nanophotonic cavity and a nearby nuclear spin, illustrated schematically in Fig.~\ref{fig:network}(a). 
The SiV is an optically active point defect in the diamond lattice \cite{hepp2014electronic,muller2014optical}. 
Its $D_{3d}$ inversion symmetry results in a vanishing electric dipole moment of the ground and excited states, rendering optical transitions insensitive to electric field noise typically present in nanofabricated structures \cite{sipahigil2014indistinguishable, sipahigil2016integrated}.
We enhance interactions between SiVs  and optical photons by incorporating them into nanocavities \cite{nguyen2019integrated}, which are critically coupled to on-chip waveguides. 
Itinerant photons in a fiber network are adiabatically transferred to this waveguide, allowing for the collection of reflected photons with efficiencies exceeding $90\%$ \cite{burek2017fiber}. 
After an initial optical characterization of the devices, a shorted, gold coplanar waveguide (CPW) is deposited in close proximity to a small subset of cavities [Fig.~\ref{fig:network} (b), inset] \cite{nguyen2019integrated}. 
This enables coherent microwave manipulation of the SiV ground state spin in a cryogenic environment ($T<\SI{0.1}{\kelvin}$), where phonon-mediated dephasing and relaxation processes are mitigated \cite{jahnke2015electron,sukachev2017silicon, pingault2017coherent}. 

%
%%%%%%%% Spin Readout %%%%%%%%%%%
%
    \begin{figure}
		\includegraphics[width=\linewidth]{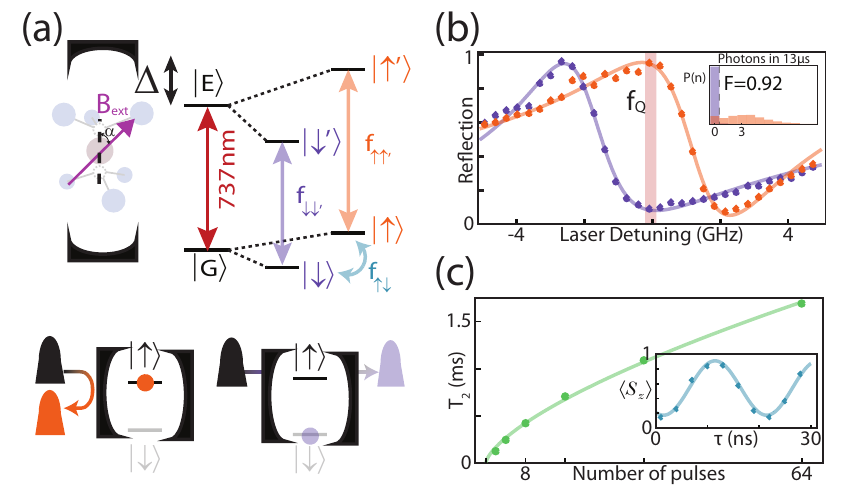}
	\caption{
			(a) Level structure of SiV spin-cavity system. The SiV optical transition at \SI{737}{\nano\meter} is coupled to the nanocavity with detuning $\Delta$. Spin conserving transitions (purple, orange) are split by an external magnetic field ($\textbf{B}_{\mathrm{ext}}$), at an angle $\alpha$ with respect to the SiV symmetry axis. Photons are only reflected by the cavity when the SiV is in state $\ket{\uparrow}$. Microwave fields at frequency $f_{\uparrow\downarrow}$ coherently drive the qubit states.
			(b) Spin-dependent reflection spectrum for $B_{ext} = 0.19$\si{\tesla}, $\alpha \approx \pi/2$ at $\Delta=0.25\kappa$. Probing at the point of maximum contrast ($f_Q$) results in high-fidelity spin-photon correlations and single-shot readout (inset, $F=0.92$).
			(c) SiV spin coherence time $T_2(N\!=\!64) >$ \SI{1.5}{\milli\second} with dynamical decoupling. (inset) Fast microwave Rabi driving of the SiV spin.
			}
	    \label{fig:cavity}
	\end{figure}
In what follows, we characterize these devices in the context of the three key ingredients of a quantum network node: (i) an efficient spin-photon interface,  (ii) a long-lived quantum memory, and (iii) access to multiple interacting qubits.

The efficient spin-photon interface is enabled by coupling to a diamond nanophotonic cavity. 
For critically-coupled cavities, the presence of an SiV modulates the bare nanocavity reflection spectrum with the strength of this modulation parametrized by the cavity cooperativity $C=4g^2/(\kappa\gamma)\sim38$ (with the single photon Rabi frequency, cavity, and atomic energy decay rate $\{g,\kappa,\gamma\}=2\pi\times\{5.6,33,0.1\}$ \si{GHz}). 
For $C>1$, we expect high-contrast modulation for a small detuning ($\Delta$) between the cavity and the SiV resonance near $737 \si{nm}$.
An external field $B_{\mathrm{ext}}$ lifts the degeneracy of the SiV spin-$\frac12$ sub-levels, creating spin-dependent reflection: photons at the frequency of maximum contrast ($f_Q$) are reflected from the cavity only when the SiV is in a specific spin state ([Fig.~\ref{fig:cavity}(a)], $\ket{\uparrow}$). 
In previous works, spin readout of the SiV was performed with $B_{\mathrm{ext}}$ parallel to the SiV symmetry axis, where the spin-conserving transitions are highly cycling \cite{sukachev2017silicon}. 
The high collection efficiency into a tapered fiber allows for fast single-shot readout of the SiV even in a misaligned field [Fig.~\ref{fig:cavity}(b)], which is necessary for the nuclear spin control described below. 
We observe a readout fidelity of $F= 0.92$ in \SI{13}{\micro\second} even when only a few ($\sim 10$) photons are scattered.
%

%%%%%%%%%%%%% MW Control %%%%%%%%%%%%%%%%%%%
We next demonstrate that the SiV spin in a nanocavity is a suitable quantum memory. 
Microwave pulses at $f_{\uparrow \downarrow} = $ \SI{6.7}{\giga\hertz} coherently manipulate the SiV spin qubit. 
The resulting Rabi oscillations, which can be driven in excess of \SI{80}{\MHz} while maintaining acceptable sample temperatures \cite{nguyen2019integrated}, are shown in the inset of Fig.~\ref{fig:cavity}(c). 
These rotations are used to probe the coherence properties of the spin via dynamical decoupling sequences [Fig.~\ref{fig:cavity}(c)] \cite{ryan2010robust,de2010universal}. 
We measure the coherence time of the SiV inside the nanocavity to be $T_2>$ \SI{1.5}{\milli\second} and scale with the number of decoupling pulses as $T_2\propto N^{2/3}$. 
The coherence scaling observed here differs from that observed in bulk diamond \cite{sukachev2017silicon}, and is similar to  NVs near surfaces \cite{Myers2014Probing}. 
This suggests that SiV memory in nanostructures is limited by an electron spin bath, for example residing near the surface of the nanostructure or resulting from implantation-induced damage \cite{nguyen2019integrated}.  

\begin{figure}
		\includegraphics[width=\linewidth]{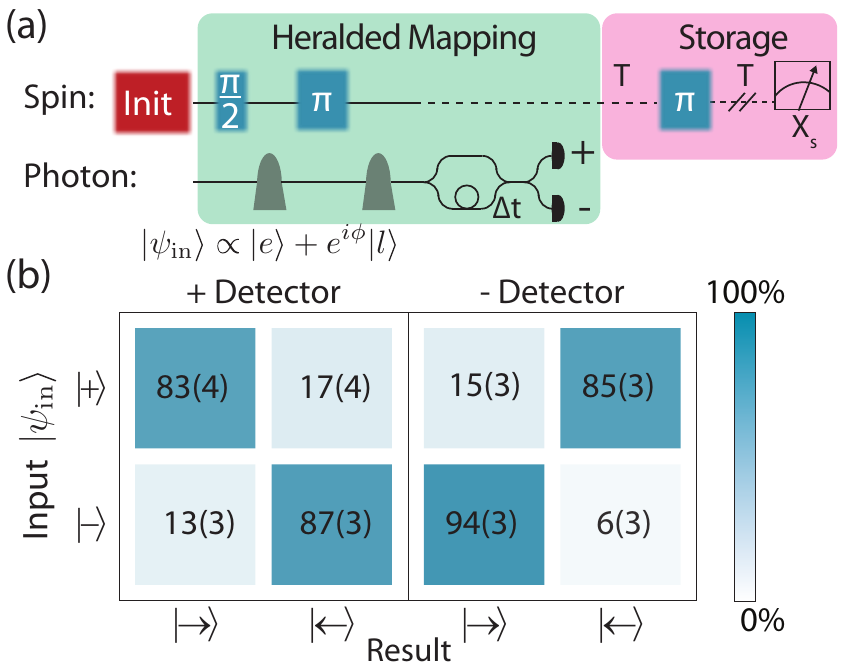}
	\caption{
			(a) Schematic for heralded photon storage. After photonic qubit is reflected off the cavity, an X measurement on the photon heralds successful state transfer which is stored for $2T = $ \SI{20}{\micro\second}. 
			(b) Spin-photon storage fidelity. The state $|\pm\rangle = \ket{\downarrow} \pm \ket{\uparrow}$ is mapped onto the SiV, with average fidelity $\mathcal{F} = 87(6)\%$.
			}
	    \label{fig:spinphoton}
	\end{figure}

%%%%%%%%%%%%%%%%%%%%% Spin-photon %%%%%%%%%%%%%%%%%%%%%%%%%%%
We now  combine the efficient spin-photon interface and control over the SiV spin state to demonstrate heralded storage of photonic qubit states in the spin-memory, a key feature of a network node \cite{kalb2015heralded}.
Fig.~\ref{fig:spinphoton}(a) outlines the experimental scheme, where photonic qubits are prepared using time-bin encoding and mapped onto the SiV spin. 
In our experiments, the SiV is first initialized into a superposition state $\ket{\rightarrow}\propto\ket{\uparrow}+\ket{\downarrow}$ by optical pumping followed by a microwave $\frac{\pi}{2}$-pulse. 
A pair of weak coherent pulses separated by $\delta t = $ \SI{30}{\nano\second} at frequency $f_Q$ are then sent to the cavity. 
The single photon sub-space corresponds to an incoming qubit state $\ket{\Psi_\mathrm{i}}\propto\beta_e\ket{e}+\beta_l\ket{l}$, where $\ket{e}$ ($\ket{l}$) denotes the presence of a photon in the early (late) time-bin. 
As a photon can only be reflected from the device if the SiV is in state $\ket{\uparrow}$ [Fig.~\ref{fig:cavity}(a)], particular components of the initial product state can be effectively "carved out" \cite{welte2017cavity}. 
We invert the SiV spin with a $\pi$-pulse between the arrival of the two time bins at the cavity, such that a photon detection event indicates that the final state has no $\ket{e\uparrow}$ or $\ket{l\downarrow}$ component. 
This leaves the system in the final spin-photon entangled state $\ket{\Psi_\mathrm{f}}\propto \beta_e\ket{e\downarrow} + \beta_l\ket{l\uparrow}$. 
The reflected photon enters a time-delay interferometer, where one arm passes through a delay line of length $\delta t$, allowing the two time-bins to interfere and erase which-time-bin information. 
As can be seen by expressing the final state in the corresponding photon basis:
\begin{equation}
\ket{\psi}_{\mathrm{f}} \propto \ket{+} (\beta_e\ket{\downarrow} + \beta_l  \ket{\uparrow}) + \ket{-} (\beta_e\ket{\downarrow} - \beta_l \ket{\uparrow}),
\end{equation}
a detection event on either the `$+$' or `$-$' arm of the interferometer represents a measurement in the X-basis ($\ket{\pm}\propto \ket{e}\pm\ket{l}$), effectively teleporting the initial photonic state onto the electron (up to a known local rotation). 
We experimentally verify generation of the entangled state $\ket{\psi}_{\mathrm{f}}$ for input states $\ket{\psi}_{\mathrm{i}} = \ket{\pm}$ by measuring spin-photon correlations \cite{nguyen2019integrated}, and use it to extract a teleportation fidelity of $0.92(6)$. 

After detection of the heralding photon, we store the teleported photonic states (initially prepared in $\{\ket{+}$ or $\ket{-}\}$) in spin memory for \SI{20}{\micro\second} by applying an additional decoupling $\pi$-pulse on the SiV spin. 
The overall fidelity of teleportation and storage is $F = 0.87(6)$ after corrected for readout errors [Fig.~\ref{fig:spinphoton}(b)]. 
The quantum storage time can be extended by additional decoupling sequences [Fig.~\ref{fig:cavity}(c)], enabling entanglement distribution up to a $T_2$-limited range of \SI{500}{\kilo\meter}. 

\begin{figure}
		\includegraphics[width=\linewidth]{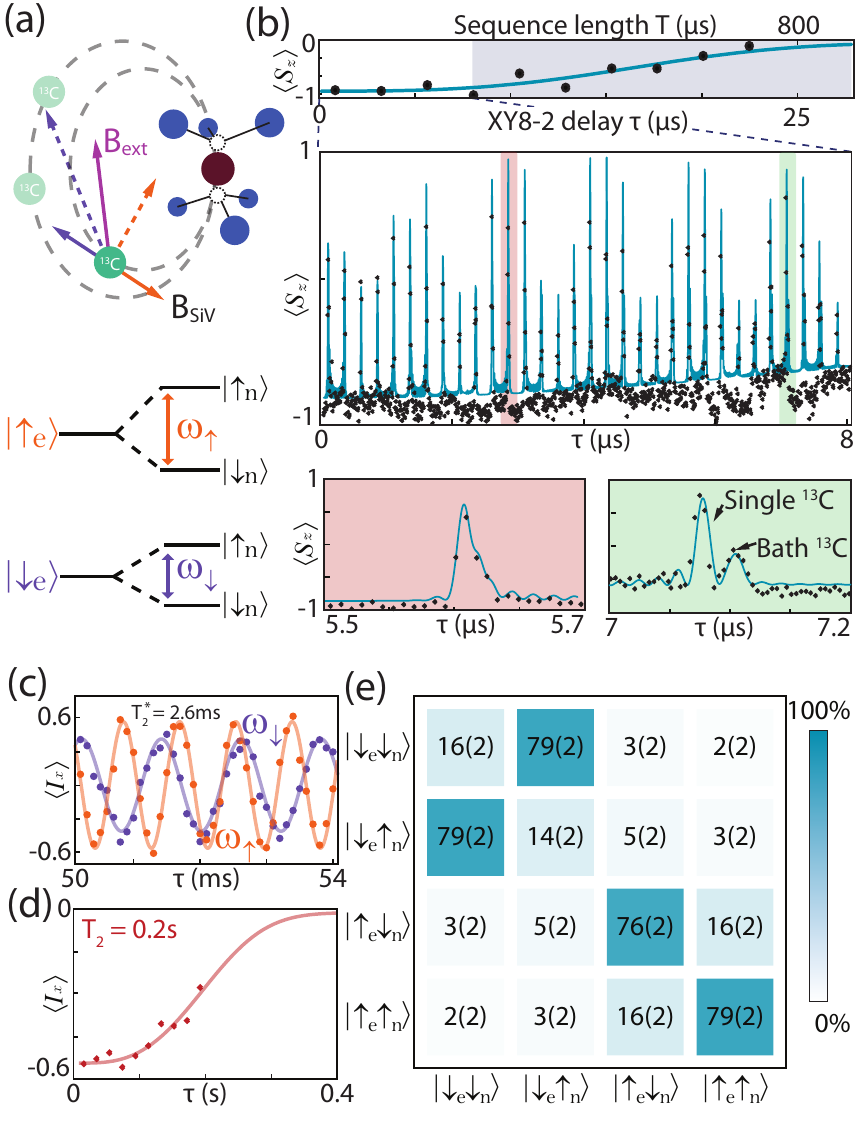}
	\caption{
			(a) Schematic of an SiV coupled to nearby \cnuc~nuclear spins. Orange (purple) vectors are conditional fields when the SiV is in state $\ket{\uparrow}$ ($\ket{\downarrow}$).
						%The larmor field of nearby nuclear spins are modulated slightly by the the spin-dependent hyperfine field produced by the SiV.
			(b) XY8-2 spin-echo. (Top) envelope for spin-echo shows a $T_2(N\!=\!16) = $\SI{603}{\micro\second}. XY8-2 at early times (Center) exhibits collapses in signal due to interaction with nuclear spins. Single \cnuc~cannot be identified at early times (red inset), but separate from the bath at long times (green inset).
			(c) Ramsey measurement on the \cnuc~nuclear spin. %Rotations of the nuclear spin are performed via composite gates on the SiV spin \cite{SOM}. 
			The nuclear spin precesses at a different Larmor frequency depending on whether the SiV is prepared in $\ket{\uparrow}$ (orange) or $\ket{\downarrow}$ (purple). Coherent oscillations persist for $T_2^* > $ \SI{2}{\milli\second} \cite{nguyen2019integrated}.
			(d) Spin echo on \cnuc, revealing $T_2 > $ \SI{0.2}{\second}
			(e) Reconstructed amplitudes for a CNOT gate transfer matrix.
			%Overall gate fidelity is $F = XX$.
			}
			\label{fig:13C}
    \end{figure}

%%%%%%%%%%%%%%%%%%%% C13 %%%%%%%%%%%%%%%%%%%%%%%%%%%
In order to extend this range and to enable more generic quantum communication protocols, we next demonstrate  a two-qubit register based on the cavity coupled SiV electronic spin and a nearby \cnuc\ nuclear memory. 
The \cnuc\ isotope of carbon is a spin-$\frac 1 2$ nucleus which has $\sim1\%$ natural abundance in diamond, and is known to exhibit exceptional coherence times \cite{bradley2019minute}. 
While direct radio-frequency manipulation of nuclear spins is impractical due to heating concerns \cite{nguyen2019integrated}, control over \cnuc\ spins can be achieved by adapting electron mediated techniques developed for Nitrogen vacancy (NV) centers \cite{dutt2007quantum,kolkowitz2012sensing,taminiau2014universal,abobeih2018one}. 
The physical principle of the SiV-\cnuc~interaction is depicted in Fig.~\ref{fig:13C}(a). 
The SiV generates a spin-dependent magnetic field $\mathbf{B}_{\mathrm{SiV}}$ at the position of the \cnuc, which is located a few lattice sites away. 
This is described by a hyperfine interaction Hamiltonian: 
\begin{equation}
\hat{H}_{\mathrm{HF}} = \hbar A_{\parallel} \frac{\hat{S}_z}{2} \frac{\hat{I}_z}{2} + \hbar A_{\perp} \frac{\hat{S}_z}{2} \frac{\hat{I}_x}{2}
\label{eq:13int}
\end{equation} 
where $\hat{S}_{z,x}$ ($\hat{I}_{z,x}$) are the Pauli operators for the electron (nuclear) spin, and $A_{\parallel, \perp}$ are the coupling parameters related to the parallel and perpendicular components of $B_{\mathrm{SiV}}$ with respect to the bias field $B_{\mathrm{ext}}$ \cite{rowan1965electron,kolkowitz2012sensing,taminiau2014universal}. 
Hyperfine interactions manifest themselves in spin-echo measurements as periodic resonances \cite{taminiau2014universal}, shown in Fig.~\ref{fig:13C}(b) for an XY8-2 decoupling sequence $\pi/2 - (\tau-\pi-\tau)^{16} - \pi/2$, where $\tau$ is the free evolution time. 
The coherence envelope for this sequence is $T_2(N=16)=\SI{603}{\micro\second}$ [Fig.~\ref{fig:13C}(b), upper panel]. 

For weakly coupled \cnuc\ ($A_\perp\ll\omega_l$, and $A_{\parallel}\ll\omega_l$, as used in this letter), the positions of the resonances \cite{taminiau2014universal}
\begin{equation}
\tau_k \approx \frac{2k+1}{2\omega_l}\left(1-\frac{1}{2}\left(\frac{A_\perp}{2 \omega_l}\right)^2\right),
\label{eq:13Cres}
\end{equation}
where $\omega_l$ is the larmor frequancy of a bare \cnuc, are insensitive to specific \cnuc\ hyperfine parameters at first order, rendering them indistinguishable at early times ($\tau_k\ll$ \SI{4}{\micro\second}, [Fig.~\ref{fig:13C}(b), red inset]). 
Individual \cnuc\ can be isolated at longer times \cite{taminiau2014universal, nguyen2019integrated}, and are used to engineer gates between a single \cnuc\ and the SiV [Fig.~\ref{fig:13C}(b), green inset]
\footnote{This is in contrast with the NV center, which is a spin-1 system and therefore features a linear shift of the resonances with coupling strength $A_\parallel$ in the $S=\{0,-1\}$ sub-system.}. 
The fundamental two-qubit gate associated with such interaction is a conditional $\pm\pi/2$ rotation of the \cnuc-spin around the $X$ axis ($R_x^{\pm\pi/2}$), which is a maximally entangling gate.
Together with unconditional rotations of the nuclear spin (which are also generated via dynamical decoupling sequences), and MW rotations on the SiV, these sequences form a universal set of gates for the register \cite{taminiau2014universal}. 

We characterize the \cnuc\ via Ramsey spectroscopy [Fig.~\ref{fig:13C}(c)]. The nuclear spin is initialized and read out via the optically addressable SiV spin by transferring population between the SiV and \cnuc\ \cite{nguyen2019integrated}. 
Depending on the SiV state before the Ramsey sequence, we observe oscillations of the nuclear spin at its eigenfrequencies $\omega_{\uparrow, \downarrow}^2 =\left( \omega_l \pm A_\parallel/2 \right)^2+\left({A_\perp}/{2} \right)^2$), allowing us to determine the hyperfine parameters $\{\omega_l, A_\parallel, A_\perp\} = 2\pi\{2.0, 0.70, -0.35\}$\si{\mega\hertz}. 
This coherence persists for $T_2^* >$ \SI{2}{\milli\second} \cite{nguyen2019integrated}, and can be further extended to $T_2 >$ \SI{0.2}{\second} by applying a single dynamical decoupling $\pi$-pulse on the nucleus, demonstrating the exceptional memory of the \cnuc\ nuclear spin [Fig.~\ref{fig:13C}(d)].

%
%%%%% CNOT and Bell%%%%%%%%%%%%%%%%%%
We benchmark the two-qubit register by demonstrating an SiV-controlled X-gate (CNOT) on the \cnuc-spin by combining a $R_x^{\pm\pi/2}$ with an unconditional nuclear $\pi/2$ rotation \cite{nguyen2019integrated}. 
This gate results in a spin flip of the \cnuc\ only if the SiV spin is in the state $\left|\downarrow\right\rangle$ [Fig.~\ref{fig:13C}(e)].
We use this gate to prepare a Bell state by initializing the register in $\left|\downarrow\downarrow\right\rangle$, and applying a $\pi/2$-rotation gate on the SiV spin followed by a CNOT gate. 
Correlation measurements yield a concurrence of $\mathcal{C}=0.22(9)$ corresponding to a Bell state fidelity of $F=0.59(4)$ after correcting for readout errors \cite{nguyen2019integrated}. 

%%%%%%%%%%%%%%% Summary  %%%%%%%%%%%%%%%%%%%%%%%
Our experiments demonstrate the first prototype of a nanophotonic quantum network node combining all necessary ingredients in a single physical system. 
We emphasize that both spin-photon and spin-spin experiments are performed in the same device under identical conditions (cavity detuning and bias field), thereby providing simultaneous demonstration of all key requirements for a network node. 

The main limitation on the demonstrated fidelities are related to the specific \cnuc\ in the proximity of the SiV, requiring an unfavorable alignment of the external magnetic field in order to isolate a single \cnuc. 
Specifically, the fidelity of two-qubit gates is limited by residual coupling to bath nuclei, SiV decoherence during the gate operations, and under/over-rotations of the nuclear spin arising from the granularity of spin-echo sequences. 
To reduce these errors, fine-tuned adaptive pulse sequences can be used to enhance sensitivity to specific nearby \cnuc, and tailor the rotation angle and axis of rotation \cite{casanova2015robust,schwartz2018robust}. 
Alternatively, replacing gold with superconducting microwave coplanar waveguides will significantly reduce ohmic heating, and allow direct radio-frequency control of nuclear spins. 
These improvements could also enable the realization of a deterministic two-qubit register based on $^\mathrm{29}$SiV, which contains both electronic and nuclear spins in a single defect \cite{nguyen2019integrated, rogers2014all}. 

The fidelity of the heralded photon storage is limited primarily by single shot readout and imperfect critical coupling of the cavity. 
The improvements of the nuclear spin control mentioned above would allow for working in an external magnetic field aligned to the SiV axis, which would improve readout fidelity from $F\sim 0.90$ (reported here) to $0.99$ \cite{evans2018photon, nguyen2019integrated}. 
The impedance mismatch of the cavity used in this experiment also gives rise to residual reflections which are not entangled with the SiV. 
Over-coupled cavities enable the use of a SiV spin-dependent phase flip for reflected photons, improving both the fidelity and success probability of spin-photon interactions.

%%%%%%%%%%%%%%%%%  Outlook %%%%%%%%%%%%%%%%%%%%%%
In conjunction with recent advances in controlling emitter inhomogeneity via electromechanical tuning \cite{machielse2019electromechanical}, these techniques should allow for chip-scale fabrication of quantum network nodes, laying the groundwork for the realization of scalable quantum repeater \cite{briegel1998repeater, childress2005fault} architectures. 
The ability to store quantum information in highly coherent \cnuc~nuclei, as well as the opportunity to extend these results to other group-IV color-centers, may open up the possibility of operating such nodes at at temperatures $>$ \SI{1}{\kelvin} \cite{metsch2019initialization,siyushev2017optical,trusheim2019lead,trusheim2018transform}. 
Finally, the efficient quantum network node demonstrated in this Letter could enable generation of multi-dimensional cluster states of many photons, which could facilitate realization of novel, ultra-fast one-way quantum communication architectures~\cite{buterakos2017deterministic}.

%%%%%%%%%%%%%% ACKNOWLEDGEMENTs %%%%%%%%%%%%%%%%%%%%%%

We would like to thank M. Markham, A. Bennett and D. Twitchen from Element Six Inc. for providing the diamond substrates used in this work, as well as F. Jelezko, R. Evans and A. Sipahigil for insightful discussions. This work was supported by DURIP grant No. N00014-15-1-28461234 through ARO, NSF, CUA, AFOSR MURI, and ARL. M.K.B. and D.S.L were supported by DoD NDSEG, B.M. and E.N.K. were supported by NSF GRFP, and R.R. was supported by the Alexander von Humboldt Foundation. Devices were fabricated at Harvard CNS, NSF award no. 1541959.

\bibliography{SiVbib}
\end{document}